\documentclass[superscriptaddress,showpacs,twocolumn]{revtex4}
\usepackage{amssymb}
\usepackage[tbtags]{amsmath}
\usepackage{graphicx}
\usepackage{epsfig,graphicx,times}

\begin{document}
\title{Coherently manipulating two-qubit quantum information using a pair of simultaneous laser pulses}
\author{L.F. Wei}
\affiliation{Frontier Research System, The Institute of Physical
and Chemical Research (RIKEN), Wako-shi, Saitama, 351-0198, Japan}
\affiliation{Institute of Quantum Optics and Quantum Information,
Department of Physics, Shanghai Jiaotong University, Shanghai
200030, P.R. China }
\author{Franco Nori}
\affiliation{Frontier Research System, The Institute of Physical
and Chemical Research (RIKEN), Wako-shi, Saitama, 351-0198, Japan}
\affiliation{Center for Theoretical Physics, Physics Department,
Center for the Study of Complex Systems, The University of
Michigan, Ann Arbor, Michigan 48109-1120}

\begin{abstract}
Several sequential operations are usually needed for implementing
controlled quantum gates and generating entanglement between a
pair of quantum bits.
Based on the conditional quantum dynamics for a two-ion system
beyond the Lamb-Dicke limit, here we propose a theoretical scheme
for manipulating two-qubit quantum information, i.e., implementing
a universal two-qubit quantum gate and generating a two-qubit
entangled state, by using a pair of simultaneous laser pulses.
Neither the Lamb-Dicke approximation nor the auxiliary atomic
level are required.
The experimental realizability of this simple approach is also
discussed.
\end{abstract}
\pacs{03.67.Lx, 32.80.Pj.} \maketitle


%
As first suggested in \cite{CZ95}, one of the most promising
scenarios for implementing a practical quantum information
processor is a system of laser-cooled trapped ions, due to its
long coherence time \cite{KS01}. In this system the qubit (i.e.,
the elementary unit of quantum information) is encoded by two
internal levels of each trapped cold ion and can be manipulated
individually by using laser pulses.
As explained below, a third auxiliary internal level of the ion is
also required.
Several key features of the proposal in \cite{CZ95}, including the
production of entangled states and the implementation of quantum
controlled operations between a pair of trapped ions, have already
been experimentally demonstrated (see, e.g.,
\cite{Monroe95,King98,Turchette98,Sackett00}).
Meanwhile, several alternative theoretical schemes (see, e.g.,
\cite{Duan01,Jonathan01,wei02,Jonathan00,Childs01,MS99-1}) have
also been developed for overcoming various difficulties in
realizing a practical ion-trap quantum information processor.
%

%
A series of laser pulses (see, e.g.,
\cite{CZ95,wei02,Jonathan00,Childs01}) and auxiliary transitions
between the encoded atomic levels and the auxiliary ones (see,
e.g., \cite{CZ95,Monroe95,Turchette98}) are usually required in
various previous schemes for manipulating two-qubit quantum
information.
For example, a {\it three}-step operation is required in
\cite{Poyatos98}, {\it five} laser pulses in \cite{CZ95,wei02},
{\it six\/} pulses in Ref. \cite{Jonathan00}, and {\it seven\/}
pulses are needed in Ref. \cite{Childs01} in order to perform a
quantum controlled-NOT (CNOT) logic gate.
The first aim of the present work is to propose an efficient
scheme for realizing quantum logic operations between a pair of
trapped cold ions by {\it a single-step operation}.
The Lamb-Dicke (LD) approximation, which requires that the
coupling between the external and internal degrees of freedom of
the ion be very weak, is made in {\it almost all} of the previous
schemes (see, e.g.,
\cite{King98,Sackett00,Jonathan01,Childs01,MS99-1}) in order to
simplify the laser-ion interaction Hamiltonian.
However, the quantum motion of the trapped ions is {\it not\/}
limited to the LD regime \cite{Wineland79}. Therefore, it is
important to manipulate quantum information stored in trapped ions
outside the LD limit.

The entanglement between different particles is a growing focus of
activity in quantum physics (see, e.g., \cite{Bennett00}), because
of experiments on non-local features of quantum mechanics and the
development of quantum information physics.
Especially, entanglement plays a central role in quantum
parallelism.
The quantum entanglement of two and four trapped cold ions have
been generated experimentally (see, e.g.,
\cite{Turchette98,Sackett00}) in the LD limit.
The second aim of the present work is to show how to
deterministically produce the entangled states of two trapped ions
outside the LD regime. The third aim is to achieve this without
auxiliary internal levels.

In summary, here we propose a scheme for manipulating quantum
information (i.e., realizing quantum controlled operations and
generating entanglement) of two trapped ions: (i) {\it beyond\/}
the LD limit, (ii) {\it without\/} needing any auxiliary atomic
level, and (iii) by using a {\it single-step operation}.
%

We consider an array of $N$ two-level cold ions of mass $M$
confined to move in the $z$ direction of a Paul trap of frequency
$\nu$.
The ions are cooled down to very low temperatures and may perform
small oscillations around their equilibrium position
$z_{i0}\,(i=1,2,...,N)$, due to their mutual repulsive Coulomb
force.
Each one of the ions is assumed to be individually addressed by a
separate laser beam.
Similarly to Ref. \cite{Solano00}, we consider the case where an
arbitrary pair (labeled by $j=1,2$) of the $N$ trapped cold ions
are illuminated independently by two weak travelling laser fields
with frequencies $\omega_j$.
%
%
The Hamiltonian corresponding to this situation is 
\begin{widetext}
\begin{eqnarray}
\hat{H}(t)=\hbar\omega_0\sum_{j=1}^{2}\frac{\hat{\sigma}_j^z}{2}
+\sum_{l=0}^{N-1}\hbar\nu_l\left(\hat{b}_l^{\dagger}\hat{b}_l+\frac{1}{2}\right)
+\frac{\hbar}{2}\sum_{j=1}^{2}\left\{\Omega_j\hat{\sigma}_j^+\exp\left[
\sum_{l=0}^{N-1}i\eta_j^l(\hat{b}_l^{\dagger}+\hat{b}_l)-i\omega_j
t-i\phi_j\right]+H.c.\right\}.
\end{eqnarray}
\end{widetext}
Here, $\nu_l\,(l=0,1,...,N-1)$ is the frequency of the $l$th mode
collective
vibrational motion,
and the LD parameter $\eta_j^l$ accounts for the coupling strength
between the internal state of the $j$th ion and the $l$th mode of
the collective vibration.
$\hat{b}_l^\dagger$ and $\hat{b}_l$ are the relevant ladder
operators.
$\Omega_j$ is the carrier Rabi frequency, which describes the
coupling strength between the laser and the $j$th ion and is
proportional to the strength of the applied laser.
$\hat{\sigma}_j^z$ and $\hat{\sigma}_j^\pm$ are Pauli operators,
$\hbar\omega_0$ is the energy separation of two internal states
$|g\rangle$ and $|e\rangle$ of the ion, and $\phi_j$ is the
initial phase of the applied laser beam.
Expanding the above Hamiltonian in terms of creation and
annihilation operators of the normal modes, we have
%
\begin{widetext}
\begin{eqnarray}
\hat{H}=\frac{\hbar }{2}\sum_{j=1,2}\left\{\Omega_{j}
\hat{\sigma}^+_j\prod_{l=0}^{N-1}\left[e^{-(\eta_j^l)^2/2}
\sum_{m,n=0}^\infty\frac{(i\eta^l_j)^{m+n}}{m\,!n!}\hat{b}_l^{\dagger
m}\hat{b}_l^n\exp[i(m-n)\nu_lt+i(\delta_jt-\phi_j)]\right]+H.c.\right\}.
\end{eqnarray}
\end{widetext}
in the interaction picture. Here, $\delta_j=\omega_j-\omega_0$ is
the detuning between the laser and the ion.
Without any loss of generality, we set the frequencies of the
applied laser beams as $\omega_j=\omega_0-k_j\nu$ with $\nu=\nu_0$
being the frequency of the center-of-mass (CM) vibrational mode,
$k_2=0$, and $k_1=1,2,...$.
Like the procedure described in \cite{Solano00, Solano01}, we
expand the above Hamiltonian in terms of creation and annihilation
operators of the normal modes and then make the usual rotating
wave approximation (RWA). For small values of $k_1$, the
excitations of the higher $l$th ($l\geq 1$) vibrational modes can
be safely neglected and the following effective Hamiltonian
\vspace{-0.2cm}
\begin{widetext}
\begin{equation}
\hat{H}=\frac{\hbar }{2}\sum_{j=1,2}\left\{ \Omega _{j}\hat{F}_j\,
\hat{\sigma}_j^+\exp\left(-\frac{\eta_j^2}{2}-i\phi_j\right)
\sum_{n=0}^\infty\frac{(i\eta_j)^{2n+k_j}\hat{a}^{\dagger
n}\hat{a}^{n+k_j}}{n!\,(n+k_j)!}+H.c.\right\}.
\end{equation}
\end{widetext}
can be obtained. Here, $\hat{a}=\hat{b}_0$,
$\hat{a}^\dagger=\hat{b}_0^\dagger$ and $\eta_j=\eta_j^0$ are the
boson operators and the LD parameter related to the CM mode,
respectively.  The operator function $\hat{F}_j=\prod_{l=1}^{N-1}
\exp[-(\eta_j^l)^2/2]
\sum_{n=0}^\infty(i\eta_j^l)^{2n}\hat{b}_l^{\dagger n}
\hat{b}_l^{n}/(n!)^2$ is irrelevant
\cite{Solano00,Solano01,Buzek02} in the weak excitation regime
($\Omega_j\ll\nu$).
Therefore, we may let $\hat{F}_j=\hat{I}$ and only label the CM
mode excitations.

It is very important to stress that, to the lowest order of the LD
parameter $\eta_j$, the effective Hamiltonian (3) reduces to that
in previous works (e.g.,
\cite{King98,Sackett00,Jonathan01,MS99-1,Solano01} under the usual
LD approximation: $(m+1/2)\eta_j^2\ll 1$.
Here, $m$ is the occupation number of the Fock state of the CM
vibrational quanta.
We now analytically solve the quantum dynamical problem associated
with the Hamiltonian (3) {\it without\/} performing the LD
approximation.
After a long derivation, we obtain the following exact
time-evolutions
%
\begin{widetext}
\begin{eqnarray}
\left\{
\begin{array}{lll}
\vspace{0.4cm} |m\rangle|g_{1}\rangle|g_{2}\rangle\rightarrow&
\cos(\tilde{\alpha}_2 t) \,
|m\rangle|g_{1}\rangle|g_{2}\rangle-ie^{-i\phi_2}\sin(\tilde{\alpha}_2 t)|m\rangle|g_{1}\rangle|e_{2}\rangle,\\
|m\rangle|g_{1}\rangle|e_{2}\rangle\rightarrow&\cos(\tilde{\alpha}_2
 \vspace{0.4cm}
t)|m\rangle|g_{1}\rangle|e_{2}\rangle-ie^{i\phi_2}\sin(\tilde{\alpha}_2
t)|m\rangle|g_{1}\rangle|g_{2}\rangle,\\
|m\rangle |e_{1}\rangle |g_{2}\rangle \rightarrow&
E_{1}(t)|m+k_1\rangle|g_{1}\rangle|g_{2}\rangle
+E_{2}(t)|m+k_1\rangle|g_{1}\rangle|e_{2}\rangle
 \vspace{0.4cm}
+E_{3}(t)|m\rangle |e_{1}\rangle|g_{2}\rangle
+E_{4}(t)|m\rangle|e_1\rangle|e_2\rangle,\\
|m\rangle|e_{1}\rangle|e_{2}\rangle\rightarrow&
F_{1}(t)|m+k_1\rangle|g_{1}\rangle|g_{2}\rangle
+F_{2}(t)|m+k_1\rangle|g_{1}\rangle|e_{2}\rangle
+F_{3}(t)|m\rangle |e_{1}\rangle |g_{2}\rangle
+F_{4}(t)\,|m\rangle |e_{1}\rangle |e_{2}\rangle,
\end{array}
\right.
\end{eqnarray}
\end{widetext}
with $k_1>m$, and 
\begin{widetext}
$$
E_1(t)=(-i)^{k_1+1}\frac{e^{i\phi_1}}{\tilde{\Delta}}\left[\sin(\tilde{\lambda}_{+}t)
-\sin(\tilde{\lambda}_{-}t)\right],\,\,
E_4(t)=-ie^{-i\phi_2}\frac{\tilde{\rho}^2}{\tilde{\Delta}}\left[
\frac{\sin(\tilde{\lambda}_{+}t)}{\tilde{\zeta}_{+}}-
\frac{\sin(\tilde{\lambda}_{-}t)}{\tilde{\zeta}_{-}}\right],
$$
 \vspace{-0.6cm}
$$
E_2(t)=(-i)^{k_1}e^{i(\phi_1-\phi_2)}
\left(\frac{\alpha_1\tilde{\rho}^2+\tilde{\gamma_2}
\tilde{\zeta}_{+}\tilde{\rho}}
{\tilde{\lambda}_{+}\tilde{\zeta}_{+}\tilde{\Delta}}\right)\left[\cos(\tilde{\lambda}_{+}t)-
\cos(\tilde{\lambda}_{-}t)\right],\,\,
E_3(t)=\frac{\tilde{\rho}^2}{\tilde{\Delta}}\left[
\frac{\cos(\tilde{\lambda}_{+}t)}{\tilde{\zeta}_{+}}
-\frac{\cos(\tilde{\lambda}_{-}t)}{\tilde{\zeta}_{-}}\right];
$$
\vspace{-0.4cm}
$$
F_1(t)=(-i)^{k_1}e^{i(\phi_1+\phi_2)}\frac{\tilde{\rho}}{\tilde{\Delta}}\left[\cos(\tilde{\lambda}_{+}t)\,
-\,\cos(\tilde{\lambda}_{-}t)\,\right],\,\,\,
F_4(t)=\frac{\tilde{\rho}^2}{\tilde{\Delta}}\,\left[
\frac{\cos(\tilde{\lambda}_{+}t)}{\tilde{\zeta}_{+}}\,
-\,\frac{\cos(\tilde{\lambda}_{-}t)}{\tilde{\zeta}_{-}}\right],
$$
\vspace{-0.4cm}
$$
F_2(t)=(-i)^{k_1+1}e^{i\phi_1}
\left(\frac{\alpha_1\tilde{\rho}^2+\tilde{\gamma}_2\tilde{\zeta}_{+}\tilde{\rho}}
{\tilde{\lambda}_{+}\tilde{\zeta}_{+}\tilde{\Delta}}\right)\left[\sin(\tilde{\lambda}_{+}t)\,
-\,\sin(\tilde{\lambda}_{-}t)\,\right],\,\,\,
F_3(t)=-ie^{i\phi_2}\frac{\tilde{\rho}^2}{\tilde{\Delta}}\left[\,
\frac{\sin(\tilde{\lambda}_{+}t)}{\tilde{\zeta}_{+}}-
\frac{\sin(\tilde{\lambda}_{-}t)}{\tilde{\zeta}_{-}}\right];
$$
\end{widetext}
Here, \vspace{-0.6cm}
\begin{widetext}
$$
\tilde{\rho}=\alpha_1\left(\tilde{\alpha}_2+\tilde{\gamma}_2\right),\,\,
\tilde{\lambda}_{\pm}=\sqrt{\frac{\tilde{\Lambda}\pm\tilde{\Delta}}{2}},\,\,
\tilde{\Lambda}=\tilde{\alpha}_2^2+\tilde{\gamma}_2^2+2\alpha_1^2,\,\,
\tilde{\Delta}^2=\tilde{\Lambda}^2-4\left(\tilde{\alpha}_2\tilde{\gamma}_2-\alpha_1^2\right)^2,\,\,
\tilde{\zeta}_{\pm}=\tilde{\lambda}^2_{\pm}-\tilde{\alpha}_2^2-\alpha_1^2;
$$
\vspace{-0.4cm}
$$
\alpha_j=\Omega^j_{m,k_j},\,
\Omega_{m,k_j}^j=\frac{\Omega_j\eta^{k_j}_je^{-\eta_j^2/2}}{2}\sqrt{\frac{(m+k_j)!}{m!}}
\sum_{n=0}^{m}\frac{(-i\eta_j)^{2n}}{(n+k_j)!}\left(
\begin{array}{c}
n\\
m \end{array} \right),\,\,
\tilde{\alpha}_j=\Omega^j_{m,0},\,
\tilde{\gamma}_j=\Omega^j_{m+k_l,0},\,j,l=1,2,\,j\neq l.
$$
\end{widetext}
%
Of course, the exact dynamical evolution for other cases can also
be derived exactly.
For example, for the case where $k_1<0$ and $|k_1|>m$\,(i.e., the
CM mode is excited by blue-sideband laser beam applied to the
first ion), the relevant evolution equations can be easily
obtained from Eq.~(4) by making the replacements
$|e_j\rangle\longleftrightarrow |g_j\rangle$ in the third and
fourth formulas.

Based on the conditional quantum dynamics derived above, we now
show how to simultaneously manipulate quantum information stored
in two ions, i.e., implementing the universal two-qubit gate and
engineering two-qubit entanglement, beyond the LD limit.
This is achieved by properly controlling the initial phases, wave
vectors, and the duration of the applied simultaneous beams.
%
%

First, the two-qubit controlled gate implies that the effect of
the operation on the second qubit (target one) depends on what
state the first qubit (control one) is in. It is easily seen from
Eq.~(4) that, if the following conditions
\begin{equation}
\cos(\tilde{\alpha}_2\tau)=\sin(\tilde{\lambda}_{+}\tau)=\sin(\tilde{\lambda}_{-}\tau)=1,
\end{equation}
are satisfied, the two-qubit controlled operation 
\vspace{-0.4cm}
\begin{widetext}
\begin{equation}
\hat{\tilde{C}}^X_{12}=|g_1\rangle\,|g_2\rangle\langle
g_1\,|\langle g_2|+|g_1\rangle\,|e_2\rangle\langle g_1\,|\langle
e_2|-ie^{-i\phi_2}|e_1\rangle\,|g_2\rangle\langle e_1\,|\langle
e_2|-ie^{i\phi_2}|e_1\rangle\,|e_2\rangle\,\langle e_1\,|\langle
g_2|,
\end{equation}
\end{widetext}
can be realized directly.
%
%
Here $\tau$ is the duration of the two simultaneously-applied
pulses. The state of the information bus is unchanged during this
operation. We notice that the controlled operation (6) implemented
here is equivalent \cite{Monroe97} to the exact CNOT gate:
$
\hat{C}^X_{12}\,=\,|g_1\rangle\,|g_2\rangle\langle g_1|\,\langle
g_2|\,+\,|g_1\rangle\,|e_2\rangle\langle g_1|\,\langle
e_2|\,+\,|e_1\rangle\,|g_2\rangle\langle e_1|\,\langle
e_2|\,+\,|e_1\rangle\,|e_2\rangle\langle e_1|\,\langle g_2|,
$
except for a local rotation. One can easily check that
both the small and large, as well as negative and positive values
of the LD parameters may be chosen to satisfy the condition (5)
for realizing the desired two-qubit controlled gate.
Our approach does not assume the LD approximation where $\eta_j\ll
1$ for $m=0$.
Thus, the present scheme operates outside the LD regime and
$\eta_j$ can be large.
%

Second, we show below that the entangled states of two trapped
ions can also be produced deterministically outside the LD regime.
In fact, it is seen from the dynamical evolution Eq.~(4) that the
entanglement between two ions can also be easily engineered
outside the LD limit.
Beginning with the non-entangled initial state
$|\psi_0\rangle=|m\rangle|g_1\rangle|g_2\rangle$, we now describe
a convenient approach to do this engineering.

We now first apply a laser beam with frequency
$\omega_1=\omega_0$, initial phase $\phi_1$, and duration $t_1$ to
ion $1$, and realize the following evolution 
\begin{widetext}
\begin{eqnarray}
|\psi_0\rangle\buildrel\hat{R}_1(m,t)\over\longrightarrow
\cos(\tilde{\alpha}_1t_1)|m\rangle|g_1\rangle|g_2\rangle-ie^{-i\phi_1}\sin(\tilde{\alpha}_1t_1)|m\rangle|e_1\rangle|g_2\rangle=|\psi_1\rangle,
\end{eqnarray}
\end{widetext}
with $\hat{R}_1(m,t_1)$ being a simple operation of rotating the
spin state of the ion $1$. The CM mode quanta is not excited and
the spin state of ion $2$ is unchanged during this process.
Obviously, this evolution may also be implemented by using a pair
of simultaneous laser beams with frequencies $\omega_1=\omega_0$
and $\omega_2=\omega_0-k_2\nu\,(k_2>m)$, respectively.

We then apply another pair of simultaneous laser beams with
frequencies $\omega_2=\omega_0$ and
$\omega_1=\omega_0-k_1\nu\,(k_1>m)$ to realize the two-qubit
controlled operation $\hat{\tilde{C}}^X_{12}$ introduced above.
This lets the non-entangled state $|\psi_1\rangle$ evolve in the
desired entangled state $|\psi_{12}(t)\rangle$
\begin{eqnarray}
|\psi_1\rangle\buildrel\hat{\tilde{C}}^X_{12}\over\longrightarrow
U(t)|g_1\rangle|g_2\rangle+V(t)|e_1\rangle|e_2\rangle=|\psi_{12}(t)\rangle,\,
\end{eqnarray}
with $
U(t_1)=\cos(\tilde{\alpha}_1t_1),\,V(t_1)=-e^{i(\phi_2-\phi_1)}\sin(\tilde{\alpha}_1t_1).
$
Interestingly, the generated entangled state reduces to the
two-qubit maximally-entangled states; i.e., the corresponding EPR
states:\,
$|\Psi^{\pm}_{12}\rangle\,=\,\left(\,|g_1\rangle\,|g_2\rangle\,
\pm\,|e_1\rangle\,|e_2\rangle\,\right)/\sqrt{2},$ if the
experimental parameters, such as the duration $t_1$ and wave
vector $\vec{\kappa}_1$ of the applied laser beam for realizing
the single-qubit rotation $\hat{R}_1(m,t_1)$, are further set
properly.



%

We now briefly discuss the experimental realizability of this
proposal. Indeed, it is not difficult to properly set the relevant
parameters for satisfying the condition (5). For example, the
desired LD parameter (see e.g., \cite{Buzek02}) defined by
%
%
\begin{equation}
\eta_j=\sqrt{\hbar \kappa_j^2/(2MN\nu)}\cos\theta_j,\,\,
\theta_j=\arccos\left(\vec{\kappa}_j\cdot\vec{z}_j/\kappa_j\right),
\end{equation}
can be reached by conveniently controlling the wave vector
$\vec{\kappa}_j$ of the applied laser beams.
It might seem at first, from the condition (5), that the present
scheme for realizing the desired gate operation cannot be easily
implemented, as the relevant experimental parameters should be set
accurately. For example, if the Rabi frequencies and LD parameters
are set as $\Omega_1=\Omega_2=\Omega,\,\eta_1=\pm
2.18403,\eta_2=\pm 1.73205$, then the duration $\tau$ of the two
applied simultaneous pulses should be set up accurately as
$\Omega\tau=56.3186$.
%
%
A simple numerical analysis shows that the lowest probability of
realizing the desired operation is still very high, even if the
relevant parameters cannot be set exactly.
For example, even if the duration of the applied laser pulses is
set roughly such that $\Omega\tau\approx 56.3$ ($56.0$), which is
$0.03\%$ ($0.57\%$) away from the exact solution of condition (5),
one can realize the operation $\hat{\tilde{C}}^X_{12}$ with a very
high probability, i.e., $99.998\%$ ($99.36\%$), via a single-step
operation.
Indeed, by testing other values we have proven that our
predictions are very robust.

Finally, we note that the duration of the applied simultaneous
pulses for realizing the above quantum controlled operation is not
much longer than that for other schemes (see, e.g.,
\cite{King98,Sackett00,Jonathan01,MS99-1}) operating in the LD
regime.
The duration for implementing the above operation is estimated
as $\thicksim 10^{-4}$ seconds, of the same order of the gate
speed operating in the LD regime \cite{Monroe95}, for
$\Omega/2\pi\,\,\approx\,\,\, 225\,\,k$Hz and
$\nu=\omega_z\,\approx\, 7\,\,M$Hz\,\cite{Turchette98}.
To excite only the chosen sidebands of the CM mode, the spectral
width of the applied laser pulse has to be small.
However, it is not so small as to affect the speed of the
operations, since the separation between the CM mode and the other
ones is sufficiently large, e.g.,
$\nu_1-\nu=(\sqrt{3}-1)\nu\sim\nu$.
Therefore, based on the current ion-trap technologies the present
scheme should be realizable in the near future.

In summary, we have proposed an efficient theoretical scheme for
simultaneously manipulating two-qubit quantum information stored
in the chosen two ions. Under certain conditions a universal
two-qubit gate can be exactly realized by a {\it single-step}
pulse process performed by simultaneously applying a pair of laser
beams with different frequencies.
By using this quantum operation, one may engineer the entanglement
state between the chosen two ions.
All operations proposed here operate {\it outside} the LD regime
and do {\it not} involve quantum transitions to auxiliary atomic
levels. It is expected that the present scheme may be extended to
simultaneously manipulate three-, four- or multi-qubit quantum
information and may be also extended for other systems besides
trapped ions, e.g., quantum dots on quantum linear supports
\cite{Brown01}, for quantum information processing.
\vspace{0.1cm}

We thank X. Hu and C. Monroe for useful comments and acknowledge
the partial support of ARDA, AFOSR, and the US National Science
Foundation grant No. EIA-0130383. 
\end{document}